\documentclass[letterpaper,english,reprint, aps]{revtex4-1}
\usepackage[T1]{fontenc}
\usepackage[latin9]{inputenc}
\usepackage{amsmath}
\usepackage{amssymb}
\usepackage{graphicx}

\makeatletter

\pdfpageheight\paperheight
\pdfpagewidth\paperwidth

\makeatother

\usepackage{babel}
\begin{document}
\preprint{APS/123-QED}
\title{Efficiency bounds for bipartite information-thermodynamic systems}
\author{Shihao Xia$^{1}$}
\author{Shuanglong Han$^{1}$}
\author{Ousi Pan$^{1}$}
\author{Yuzhuo Pan$^{2}$}
\author{Jincan Chen$^{1}$}
\author{Shanhe Su$^{1}$}
\email{sushanhe@xmu.edu.cn}

\address{$^{1}$Department of Physics, Xiamen University, Xiamen 361005, People's
Republic of China~\\
$^{2}$College of Physics and Information Engineering, Quanzhou Normal
University, 362000, People's Republic of China}
\date{\today}
\begin{abstract}
This study introduces a novel approach to derive a lower bound for
the entropy production rate of a subsystem by utilizing the Cauchy-Schwarz
inequality. It extends to establishing comprehensive upper and lower
bounds for the efficiency of two subsystems. These bounds are applicable
to a wide range of Markovian stochastic processes, which enhances
the accuracy in depicting the range of energy conversion efficiency
between subsystems. Empirical validation is conducted using a two-quantum-dot
system model, which serves to confirm the effectiveness of our inequality
in refining the boundaries of efficiency. 
\end{abstract}
\maketitle

\section{introduction}

Over the past two decades, the field of stochastic thermodynamics
has witnessed remarkable progress, greatly enriching our comprehension
of physical phenomena occurring in small fluctuating systems. This
framework has unveiled the inherent connection between thermodynamic
irreversibility and non-equilibrium fluctuations. 

Simultaneously, the study of stochastic thermodynamics has facilitated
explorations into the interrelationship between information and thermodynamics,
as well as the thermodynamic limits of fluctuating mesoscopic systems.
\citep{seifert2012stochastic,van2015ensemble}. A pivotal discovery
in this domain is Landauer's principle \citep{landauer1961irreversibility},
which quantifies the heat dissipation required for the erasure of
information in the limit of slow quasi-static processes \citep{dong2011quantum}.
This principle has been empirically validated in both classical and
quantum frameworks \citep{berut2012experimental,dago2021information,hashimoto2022effect}.
In the era of high-speed computational demands, a noteworthy breakthrough
has been the investigation of the energy consumption associated with
finite-time information erasure \citep{van2022finite,proesmans2020finite,zhen2021universal}.
Leveraging the unique synergy between information and thermodynamics,
researchers have devised various heat engines, such as the Szilard
engine \citep{kim2011quantum,koski2014experimental,malgaretti2022szilard}
and molecular engines \citep{chen2017molecular,leighton2023inferring,barato2015thermodynamic}.
These advancements have opened up new avenues for the design of miniature
heat engines. 

On the other hand, thermodynamic inequalities have established foundational
limits on entropy production and energy usage \citep{barato2015thermodynamic,horowitz2020thermodynamic,shiraishi2021optimal},
deepening our understanding of thermodynamics and providing valuable
insights into microsystems. A multitude of studies have corroborated
the Clausius statement of the second law of thermodynamics by employing
thermodynamic inequalities. This consensus and the substantial research
interests in this field signify its importance and widespread relevance.
\citep{shiraishi2016universal,pietzonka2018universal,funo2019speed,shiraishi2018speed,van2021geometrical,aurell2011optimal,van2021lower,tajima2021superconducting,barato2015thermodynamic,crooks2007measuring,salamon1983thermodynamic,sivak2012thermodynamic,scandi2019thermodynamic,chen2021extrapolating}.
In the pursuit of establishing theoretical bounds, researchers have
adopted various methodologies. Notably, Jensen bounds \citep{leighton2023inferring,leighton2022dynamic}
and Cramér-Rao \citep{tanogami2023universal} bounds represent significant
approaches in delineating these constraints. 

While existing researchs provide valuable insights into the boundaries
of energy conversion and entropy production, our investigation adopts
an alternative framework. Specifically, we leverage the Cauchy-Schwarz
inequality \citep{funo2019speed,shiraishi2018speed} within the stochastic
thermodynamics context to derive a more refined lower bound for the
entropy production rate in subsystems. This approach enables us to
obtain a accurate approximation to the energy conversion efficiency
of these subsystems.

The structure of this paper is organized as follows: Section \ref{sec:2}
delves into the examination of stochastic thermodynamics in bipartite
systems, where the efficiency bounds for subsystems are derived. Section
\ref{sec:3} focuses on the double quantum dot system and derives
expressions for the associated thermodynamic quantities by calculating
the transition frequency. By building upon the double quantum dot
model, section \ref{sec:4} presents our findings in a detailed discussion.
Finally, the paper concludes with section \ref{sec:5}, which summarizes
the key insights gained from the study.

\section{Stochastic thermodynamics of bipartite systems\label{sec:2}}

\subsection{Stochastic dynamics for a bipartite system}

We adopt a conventional stochastic process model for a bipartite system
$Z$, which is described thoroughly and extensively in \citep{horowitz2014thermodynamics}.
The state of $Z$ is described by $z=(x,y)$ with $x$ and $y$ representing
the particle counts in subsystems $X$ and $Y$, respectively. The
state $z$'s time-dependent probability, $p(z)$, evolves following
the Markovian master equation

\begin{equation}
\dot{p}(z)=\sum_{z^{\prime},v}J_{zz^{\prime}}^{v}.\label{eq:mastereq}
\end{equation}

Here, the flux $J_{zz^{\prime}}^{v}=W_{zz^{\prime}}^{v}p\left(z^{\prime}\right)-W_{z^{\prime}z}^{v}p(z)$
hinges on the time-independent transition rate $W_{zz^{\prime}}^{v}$
from state $z^{\prime}$ to state $z$ induced by reservoir $v$,
and the reverse transition rate $W_{z^{\prime}z}^{v}$. The transition
rate matrix, being non-negative $W_{zz^{\prime}}^{v}\geq0$ for $z\neq z^{\prime}$,
restricts the bipartite structure from allowing simultaneous state
changes in both subsystems. Consequently, the form of $W_{zz^{\prime}}^{v}$
is constrained to

\begin{equation}
W_{zz^{\prime}}^{v}=\begin{cases}
W_{xx^{\prime}\mid y}^{v} & \left(x\neq x^{\prime},y=y^{\prime}\right)\\
W_{yy^{\prime}\mid x}^{v} & \left(x=x^{\prime},y\neq y^{\prime}\right)\\
0, & \text{ (otherwise) }
\end{cases}
\end{equation}
where $W_{xx^{\prime}\mid y}^{v}$ signifies the transition rate for
subsystem $X$ moving from state $x^{\prime}$ to state $x$ given
that subsystem $Y$ is in state $y$, and similarly, $W_{yy^{\prime}\mid x}^{v}$
for transitions within subsystem $Y$.

\subsection{Information thermodynamics of the bipartite system}

Due to the bipartite nature, the entropy production rate $\dot{\sigma}$
of the bipartite system can be split into two components:
\begin{equation}
\dot{\sigma}=\dot{\sigma}^{X}+\dot{\sigma}^{Y}.
\end{equation}
Herein, $\dot{\sigma}^{X}$ and $\dot{\sigma}^{Y}$ represent the
partial entropy production rates resulting from transitions in $X$
and $Y$, respectively, which are expressed as

\begin{equation}
\dot{\sigma}^{X}=\frac{1}{2}\sum_{x,x^{\prime},y,v}J_{xx^{\prime}\mid y}^{v}\ln\frac{p\left(x^{\prime},y\right)W_{xx^{\prime}\mid y}^{v}}{p(x,y)W_{x^{\prime}x\mid y}^{v}}\label{eq:EPX}
\end{equation}
and

\begin{equation}
\dot{\sigma}^{Y}=\frac{1}{2}\sum_{y,y^{\prime},x,v}J_{yy^{\prime}\mid x}^{v}\ln\frac{p\left(y^{\prime},x\right)W_{yy^{\prime}\mid x}^{v}}{p(y,x)W_{y^{\prime}y\mid x}^{v}}.\label{eq:EPY}
\end{equation}

The entropy flow towards the environment, $\dot{S}_{r}$, is given
by:

\begin{equation}
\dot{S}_{r}=\dot{S}_{r}^{X}+\dot{S}_{r}^{Y}
\end{equation}
where 

\begin{equation}
\dot{S}_{r}^{X}=\frac{1}{2}\sum_{x,x^{\prime},y,v}J_{xx^{\prime}\mid y}^{v}\ln\frac{W_{xx^{\prime}\mid y}^{v}}{W_{x^{\prime}x\mid y}^{v}}\label{eq:SRX}
\end{equation}
and

\begin{equation}
\dot{S}_{r}^{Y}=\frac{1}{2}\sum_{x,x^{\prime},y,v}J_{yy^{\prime}\mid x}^{v}\ln\frac{W_{yy^{\prime}\mid x}^{v}}{W_{y^{\prime}y\mid x}^{v}},\label{eq:SRY}
\end{equation}
correspond to the energy flux from $X$ and $Y$ to their respective
environment. 

\begin{equation}
\dot{I}^{X}=\frac{1}{2}\sum_{x,x^{\prime},y,v}J_{xx^{\prime}\mid y}^{v}\ln\frac{p(y\mid x)}{p\left(y\mid x^{\prime}\right)}\label{eq:IX}
\end{equation}
and

\begin{equation}
\dot{I}^{Y}=\frac{1}{2}\sum_{x,x^{\prime},y,v}J_{yy^{\prime}\mid x}^{v}\ln\frac{p(x\mid y)}{p\left(x\mid y^{\prime}\right)}\label{eq:IY}
\end{equation}
quantify how information exchanges between the two subsystems. When
$\dot{I}^{X}>0$, subsystem $X$ is measuring $Y$; vice versa, $\dot{I}^{X}<0$
signifies that $X$ is consuming information in order to extract energy.

From Eq. (\ref{eq:EPX}) to Eq. (\ref{eq:IY}), it is evident that
$\dot{\sigma}^{X}$ and $\dot{\sigma}^{Y}$ must be non-negative,
which can be split into three components as \citep{su2022theoretical}:

\begin{equation}
\dot{\sigma}^{X}=\dot{S}^{X}+\dot{S_{r}}^{X}-\dot{I}^{X}\geq0\label{eq:Tx}
\end{equation}

\begin{equation}
\dot{\sigma}^{Y}=\dot{S}^{Y}+\dot{S_{r}}^{Y}-\dot{I}^{Y}\geq0.\label{eq:Ty}
\end{equation}

Upon reaching a steady state, the system settles into a stationary
probability distribution $[\dot{p}(x)=0\mathrm{\,and}\dot{p}(y)=0]$,
leading to $\dot{S}^{X(Y)}=0$. Thus, Eqs. (\ref{eq:Tx}) and (\ref{eq:Ty})
are simplified as follows:

\begin{equation}
\dot{\sigma}^{X}=\dot{S}_{r}^{X}-\dot{I}^{X}\geq0\label{eq:T2x}
\end{equation}

and

\begin{equation}
\dot{\sigma}^{Y}=\dot{S_{r}}^{Y}-\dot{I}^{Y}\geq0.\label{eq:T2y}
\end{equation}

In scenarios where $\dot{I}^{X}>0$, as Eq. (\ref{eq:T2x}), the information
$\dot{I}^{X}$ is constrained by the entropy flow to the environment
$\dot{S}_{r}{}^{X}$, which represents the dissipation required for
$X$ to learn about $Y$. Information then becomes a resource for
energy extraction or work execution via feedback on $Y$. 

\subsection{The thermodynamic efficiency of two subsystems}

We now shift our focus to the nonequilibrium steady state, in which
each subsystem exchanges work and heat with their respective external
reservoirs. To articulate thermodynamic quantities, it's assumed that
the transition rates adhere to the local detailed balance condition
specified by the following equation\citep{peliti2021stochastic,tanogami2023universal}

\begin{equation}
\ln\frac{W_{xx^{\prime}\mid y}^{v}}{W_{x^{\prime}x\mid y}^{v}}=\beta\left(\epsilon_{x^{\prime}y}-\epsilon_{xy}+\Delta_{xx^{\prime}\mid y}^{v}\right),\label{eq:detailedbalance}
\end{equation}
where $\beta=\left(k_{\mathrm{B}}T\right)^{-1}$ is the inverse temperature
by assuming that all the reservoirs are at temperature $T$, $\epsilon_{xy}$
represents the energy of state $(x,y)$, and $\Delta_{xx^{\prime}}^{y}$
is the energy provided by an external agent during the transition
$\left(x^{\prime},y\right)\rightarrow(x,y)$. The parameter $\Delta_{xx^{\prime}}^{y}$
will affect the heat flow. For the quantum dot system interacting
with an electron reservoir, $-\Delta_{xx^{\prime}}^{y}$ represents
the chemical potential of the reservoir, which will be further illustrated
in Sec. III. The average rate of heat absorbed by $X$ from the environment
is then determined by

\begin{equation}
\dot{Q}^{X}=-\frac{k_{\mathrm{B}}T}{2}\sum_{x,x^{\prime},y,v}J_{xx^{\prime}\mid y}^{v}\ln\frac{W_{xx^{\prime}\mid y}^{v}}{W_{x^{\prime}x\mid y}^{v}}.
\end{equation}

The average rate of work done by the external agent on $X$ is identified
as

\begin{equation}
\dot{W}^{X}=\frac{1}{2}\sum_{x,x^{\prime},y,v}J_{xx^{\prime}\mid y}^{v}\Delta_{xx^{\prime}\mid y}^{v}.
\end{equation}

Subsequently, the average rate of change of internal energy, interpreted
as the work done by $X$ on $Y$, is expressed as

\begin{equation}
\dot{W}^{X\rightarrow Y}=\frac{1}{2}\sum_{x,x^{\prime},y,v}J_{xx^{\prime}\mid y}^{v}\left(\epsilon_{xy}-\epsilon_{x^{\prime}y}\right).
\end{equation}
The rates $\dot{W}^{Y}$, $\dot{Q}^{Y}$, and $\dot{W}^{Y\rightarrow X}$
are similarly defined for $Y$. Note that each subsystem adheres to
the first law of thermodynamics that encapsulates local energy conservation:

\begin{equation}
\begin{aligned}\dot{W}^{X\rightarrow Y} & =\dot{W}^{X}+\dot{Q}^{X},\\
\dot{W}^{Y\rightarrow X} & =\dot{W}^{Y}+\dot{Q}^{Y}.
\end{aligned}
\label{eq:WXY}
\end{equation}
Furthermore, at steady state, the powers and information flows satisfy
$\dot{W}^{X\rightarrow Y}+\dot{W}^{Y\rightarrow X}=0$ and $\dot{I}^{X}+\dot{I}^{Y}=0$.
On the basis of Eqs. (\ref{eq:WXY}), (\ref{eq:T2x}), and (\ref{eq:T2y}),
each subsystem conforms to the second law of thermodynamics at the
steady state:

\begin{equation}
\begin{aligned}\dot{\sigma}^{X} & =\beta\dot{W}^{X}-\beta\dot{W}^{X\rightarrow Y}-\dot{I}^{X}\geq0,\\
\dot{\sigma}^{Y} & =\beta\dot{W}^{Y}-\beta\dot{W}^{Y\rightarrow X}-\dot{I}^{Y}\geq0.
\end{aligned}
\label{eq:T2XY}
\end{equation}

Within this framework, each subsystem satisfies the first and second
laws. Without loss of generality, let $Y$ be the \textquotedbl upstream\textquotedbl{}
subsystem, with the rate of work input to $Y$ being positive, denoted
as $\dot{W}^{Y}>0$. We will limit our discussion to functional machines
that produce work from subsystem $X$, i.e., $\dot{W}^{X}\leq0$.
They thus each have their own thermodynamic efficiencies:
\begin{equation}
\begin{aligned}\eta^{Y} & :=\frac{\beta\dot{W}^{X\rightarrow Y}+\dot{I}^{Y}}{\beta\dot{W}^{Y}},\\
\eta^{X} & :=\frac{-\beta\dot{W}^{X}}{\beta\dot{W}^{Y\rightarrow X}+\dot{I}^{Y}}.
\end{aligned}
\end{equation}

As discussed in Ref. \citep{leighton2023inferring} , the efficiency
of subsystem $Y$, denoted as $\eta^{Y}$, reflects how effectively
$Y$ transforms the input work into free energy available for subsystem
$X$. Conversely, $\eta^{X}$ measures the efficiency of $X$ in converting
this free energy back into output work. The product of these efficiencies,
$\eta^{Y}\eta^{X}=-\dot{W}^{X}/\dot{W}^{Y}=\eta^{T}$, defines the
overall thermodynamic efficiency of the system. These efficiencies
remain valid as long as both $\dot{W}_{Y}$ and $\beta\dot{W}_{Y\rightarrow X}+\dot{I}_{Y}$
are strictly positive. This ensures that the conditions $0\leq\eta^{X}\leq1$
and $0\leq\eta^{Y}\leq1$ are satisfied, as derived from Eq. (\ref{eq:T2XY}). 

\subsection{Bounds on the efficiencies of two subsystems}

Equation (\ref{eq:T2XY}) yields dual equalities:
\begin{equation}
\dot{\sigma}^{X}-\beta\dot{W}^{X}=\beta\dot{W}^{Y\rightarrow X}+\dot{I}^{Y}=\beta\dot{W}^{Y}-\dot{\sigma}^{Y}.
\end{equation}

Applying any lower bounds $\dot{\sigma}_{LB}^{X}\leq\dot{\sigma}^{X}$
and $\dot{\sigma}_{LB}^{Y}\leq\dot{\sigma}^{Y}$ to this equation
facilitates the derivation of upper and lower bounds on $\beta\dot{W}^{Y\rightarrow X}+\dot{I}^{Y}$:
\begin{equation}
\dot{\sigma}_{LB}^{X}-\beta\dot{W}^{X}\leq\beta\dot{W}^{Y\rightarrow X}+\dot{I}^{Y}\leq\beta\dot{W}^{Y}-\dot{\sigma}_{LB}^{Y}.\label{eq:inequality}
\end{equation}

Dividing Eq. (\ref{eq:inequality}) by $\dot{W}^{Y}$ results in upper
and lower bounds on the efficiency of $Y$:
\begin{equation}
\eta^{T}\left(1+\frac{\dot{\sigma}_{LB}^{X}}{-\beta\dot{W}^{X}}\right)\leq\eta^{Y}\leq1-\frac{\dot{\sigma}_{LB}^{Y}}{\beta\dot{W}^{Y}}.\label{eq:mainY}
\end{equation}

Conversely, multiplying the reciprocal of Eq. (\ref{eq:inequality})
by $-\dot{W}^{X}$ produces upper and lower bounds on the efficiency
of $X$:

\begin{equation}
\eta^{T}\left(1-\frac{\dot{\sigma}_{LB}^{Y}}{\beta\dot{W}^{Y}}\right)^{-1}\leq\eta^{X}\leq\left(1+\frac{\dot{\sigma}_{LB}^{X}}{-\beta\dot{W}^{X}}\right)^{-1}.\label{eq:mainX}
\end{equation}

In the context of Markovian dynamics, the recently derived bound \citep{su2022theoretical}
provides tighter lower bounds of the partial entropy production rates
$\dot{\sigma}^{X}$ and $\dot{\sigma}^{Y}$ for the Markovian dynamics
{[}also see Appendix. \ref{sec:appendix}{]}:

\begin{equation}
\begin{aligned}\dot{\sigma}^{X} & \geq\frac{\left(\dot{S}_{r}^{X}\right)^{2}}{\Theta^{X}},\\
\dot{\sigma}^{Y} & \geq\frac{\left(\dot{S}_{r}^{Y}\right)^{2}}{\Theta^{Y}},
\end{aligned}
\label{eq:epbounds}
\end{equation}
where

\begin{equation}
\ensuremath{\Theta^{X}=\frac{1}{2}\sum_{x,x^{\prime},y,v}\left(\ln\frac{W_{xx^{\prime}\mid y}^{v}}{W_{x^{\prime}x\mid y}^{v}}\right)^{2}W_{xx^{\prime}\mid y}^{v}p\left(x^{\prime},y\right)},\label{eq:THX}
\end{equation}
and

\begin{equation}
\ensuremath{\Theta^{Y}=\frac{1}{2}\sum_{y,y^{\prime},x,v}\left(\ln\frac{W_{yy^{\prime}\mid x}^{v}}{W_{y^{\prime}y\mid x}^{v}}\right)^{2}W_{yy^{\prime}\mid x}^{v}p\left(y^{\prime},x\right)}.\label{eq:THY}
\end{equation}

The inequalities specified in Eq. (\ref{eq:epbounds}) clearly articulate
the dynamic behaviors within a system interacting with a thermal bath.
When bath $v$ triggers a transition from state $\left(x^{\prime},y\right)$
to $(x,y)$, according to Eq. (\ref{eq:detailedbalance}), the energy
absorbed by the system during this process is represented by the equation
$E(x,y)-E\left(x^{\prime},y\right)=-k_{B}T_{v}\ln\frac{W_{xx^{\prime}\mid y}^{v}}{W_{x^{\prime}x\mid y}^{v}}$.
Here, $E(x,y)$ denotes the energy of the state $(x,y)$, $k_{B}$
is the Boltzmann constant, and $T_{v}$ represents the temperature
of bath $v$. The transition probability coefficient, $W_{xx^{\prime}\mid y}^{v}p\left(x^{\prime},y\right)$,
measures the rate at which transitions between these states occur.
Additionally, the factor $\Theta^{X/Y}$ can be understood as a dynamic
activity factor, quantifying the level of active transitions influenced
by the energy exchanges between the states and the bath. Inserting
these bounds into Eqs. (\ref{eq:mainX}) and (\ref{eq:mainY}) gives
\begin{equation}
\begin{gathered}\eta^{T}\left[1+\frac{\left(\dot{S}_{r}^{X}\right)^{2}}{-\Theta^{X}\dot{W}^{X}}\right]\leq\eta^{Y}\leq1-\frac{\left(\dot{S}_{r}^{Y}\right)^{2}}{\Theta^{Y}\dot{W}^{Y}},\\
\eta^{T}\left[1-\frac{\left(\dot{S}_{r}^{Y}\right)^{2}}{\Theta^{Y}\dot{W}^{Y}}\right]^{-1}\leq\eta^{X}\leq\left[1+\frac{\left(\dot{S}_{r}^{X}\right)^{2}}{-\Theta^{X}\dot{W}^{X}}\right]^{-1}.
\end{gathered}
\label{eq:effbound}
\end{equation}

 These two inequalities encapsulate a thorough formulation of our
principal results, providing a systematic method for establishing
efficiency boundaries for subsystems in relation to their interactions
with the environment and determining minimal rates of entropy production.
Then we define $\eta_{L}^{Y}=\eta^{T}\left[1+\frac{\left(\dot{S}_{r}^{X}\right)^{2}}{-\Theta^{X}\dot{W}^{X}}\right]$,
$\eta_{U}^{Y}=1-\frac{\left(\dot{S}_{r}^{Y}\right)^{2}}{\Theta^{Y}\dot{W}^{Y}}$,
$\eta_{L}^{X}=\eta^{T}\left[1-\frac{\left(\dot{S}_{r}^{Y}\right)^{2}}{\Theta^{Y}\dot{W}^{Y}}\right]^{-1}$
and $\eta_{U}^{X}=\left[1+\frac{\left(\dot{S}_{r}^{X}\right)^{2}}{-\Theta^{X}\dot{W}^{X}}\right]^{-1}$.
As noted, Eq. (29) in our work bears similarity to Eq. (18) from
Ref. \citep{leighton2023inferring}. This is because we have employed
the same procedure to determine the bound on the efficiency of the
two subsystems. Specifically, this involves substituting the explicit
expression for the lower bound of the entropy production rate. However,
owing to the differing system descriptions used, we have employed
distinct approaches to derive the lower bound of the entropy production
rate. In Ref. \citep{leighton2023inferring}, the model is characterized
by the overdamped Langevin equation, and the lower bound on the entropy
production rate was derived using Jensen's inequality as detailed
in \citep{leighton2022dynamic}. Our findings are anchored within
the framework of stochastic thermodynamics. For establishing the lower
bound of the entropy production rate, we utilized the Cauchy-Schwarz
inequality. Unlike Jensen's inequality, which necessitates convexity
of the function, Cauchy-Schwarz inequality imposes no such restriction.
This property renders Cauchy-Schwarz inequality more versatile within
the context of stochastic thermodynamics.

\section{The double-quantum-dot system\label{sec:3}}

\begin{figure}
\includegraphics[scale=0.4]{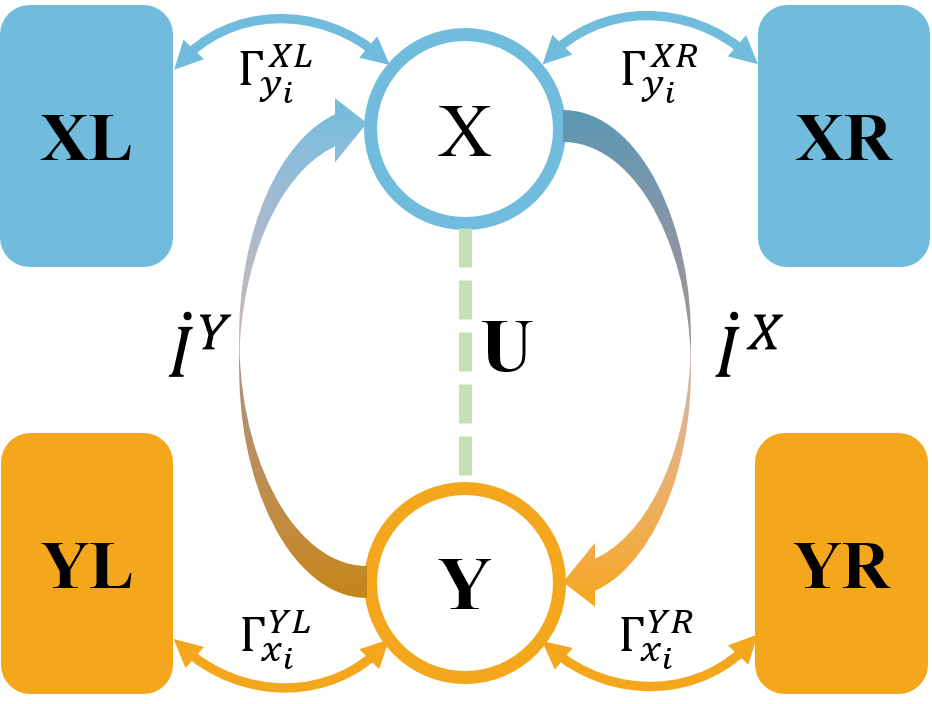}\caption{The schematic diagram of a double-quantum-dot system}
\end{figure}

The double-quantum-dot (DQD) system consists of two quantum dots interacting
via a Coulomb repulsion $U$. The DQD {[}See Fig. 1{]} is modeled
by the Hamiltonian 
\begin{equation}
H_{S}=\varepsilon_{X}d_{X}^{\dagger}d_{X}+\varepsilon_{Y}d_{Y}^{\dagger}d_{Y}+Ud_{X}^{\dagger}d_{X}d_{Y}^{\dagger}d_{Y},\label{eq:Hs}
\end{equation}
where $d_{X}^{\dagger}\left(d_{X}\right)$ creates (annihilates) one
electron on QD $X$ with energy $\varepsilon_{X}$, and $d_{Y}^{\dagger}\left(d_{Y}\right)$
creates (annihilates) one electron on QD $Y$ with energy $\varepsilon_{Y}$.
QD $X(Y)$ is weakly coupled to two Fermi reservoirs $XL$ and $XR$
($YL$ and $YR$). Thus, electrons are transported through parallel
interacting channels. The Hamiltonian of the reservoirs is given by
\begin{equation}
H_{B}=\sum_{k}\sum_{\alpha\in\{X,Y\}}\sum_{\beta\in\{L,R\}}\varepsilon_{k\alpha\beta}c_{k\alpha\beta}^{\dagger}c_{k\alpha\beta},\label{eq:Hb}
\end{equation}
where $c_{k\alpha\beta}^{\dagger}\left(c_{k\alpha\beta}\right)$ is
the creation (annihilation) operator at energy level $\varepsilon_{k\alpha\beta}$
in bath $\alpha\beta$. The interaction between the DQD and the environment
reads 
\begin{equation}
H_{I}=\sum_{k}\sum_{\alpha\in\{X,Y\}}\sum_{\beta\in\{L,R\}}\left(t_{k\alpha\beta}d_{\alpha}c_{k\alpha\beta}^{\dagger}+t_{k\alpha\beta}^{*}c_{k\alpha\beta}d_{\alpha}^{\dagger}\right),\label{eq:Hi}
\end{equation}
where $t_{k\alpha\beta}$ denotes the coupling strength of the transition
between QD $\alpha$ and reservoir $\alpha\beta$ at energy level
$\varepsilon_{k\alpha\beta}$. We use $x_{0}$ and $y_{0}\left(x_{1}\right.$
and $\left.y_{1}\right)$ to denote that $X$ and $Y$ are in empty
(filled) states, respectively. The energy eigenstates of the DQD coincide
with the localized Fock states $\left|\left(x_{0},y_{0}\right)\right\rangle ,\left|\left(x_{1},y_{0}\right)\right\rangle ,\left|\left(x_{0},y_{1}\right)\right\rangle $,
and $\left|\left(x_{1},y_{1}\right)\right\rangle $, where their respective
eigenvalues are $0,\varepsilon_{X},\varepsilon_{Y}$, and $\varepsilon_{X}+\varepsilon_{Y}+U$.

For a non-degenerate system weakly coupled to different environmental
modes, the dynamics of the populations satisfies Eq. (\ref{eq:mastereq}).
The transition rates follow from Fermi's golden rule and are given
by 
\[
\begin{aligned} & W_{x_{1}x_{0}\mid y_{i}}^{v}=\Gamma_{y_{i}}^{v}f_{y_{i}}^{v},\\
 & W_{x_{0}x_{1}\mid y_{i}}^{v}=\Gamma_{y_{i}}^{v}\left(1-f_{y_{i}}^{v}\right),\\
 & W_{y_{1}y_{0}\mid x_{i}}^{v}=\Gamma_{x_{i}}^{v}f_{x_{i}}^{v},\\
 & W_{y_{0}y_{1}\mid x_{i}}^{v}=\Gamma_{x_{i}}^{v}\left(1-f_{x_{i}}^{v}\right),
\end{aligned}
\]
where $f_{x_{i}}^{v}=\left\{ 1+\exp\left[\beta_{v}\left(\varepsilon_{Y}+iU-\mu_{v}\right)\right]\right\} ^{-1}$
and $f_{y_{i}}^{v}=\left\{ 1+\exp\left[\beta_{v}\left(\varepsilon_{X}+iU-\mu_{v}\right)\right]\right\} ^{-1}(i=0,1)$
are the Fermi distribution functions, $\beta_{v}=1/\left(k_{\mathrm{B}}T_{v}\right)$,
and $\Gamma_{y_{i}}^{v}\left(\Gamma_{x_{i}}^{v}\right)$ is a positive
constant that characterizes the height of the potential barrier between
$X(Y)$ and reservoir $v$. The potential barrier of $Y$, characterized
by $\Gamma_{x_{i}}^{v}$, depends on the state of $X$, and vice versa.
Note that $\ln\frac{W_{x_{1}x_{0}\mid y_{0}}^{v}}{W_{x_{0}x_{1}\mid y_{0}}^{v}}=\beta_{v}\left(\varepsilon_{X}-\mu_{v}\right)$
and $\ln\frac{W_{x_{1}x_{0}\mid y_{1}}^{v}}{W_{x_{0}x_{1}\mid y_{1}}^{v}}=\beta_{v}\left(\varepsilon_{X}+U-\mu_{v}\right)$.
Thus, for the quantum dot model, the parameter $\Delta_{xx^{\prime}}^{y}$
in Eq. (\ref{eq:detailedbalance}) is equal to $-\mu_{v}$ and will
affect the heat flow. 

For the purpose of reducing the number of parameters, the energy dependences
of the tunneling rates are parametrized by dimensionless parameters
$\delta$ and $\Delta$, i.e., 
\begin{equation}
\begin{aligned}\Gamma_{x_{0}}^{YL} & =\Gamma\frac{e^{\Delta}}{\cosh(\Delta)}, & \Gamma_{x_{1}}^{YL}=\Gamma\frac{e^{-\Delta}}{\cosh(\Delta)},\\
\Gamma_{x_{0}}^{YR} & =\Gamma\frac{e^{-\Delta}}{\cosh(\Delta)}, & \Gamma_{x_{1}}^{YR}=\Gamma\frac{e^{\Delta}}{\cosh(\Delta)},\\
\Gamma_{y_{0}}^{XL} & =\Gamma\frac{e^{\delta}}{\cosh(\delta)}, & \Gamma_{y_{1}}^{XL}=\Gamma\frac{e^{-\delta}}{\cosh(\delta)},\\
\Gamma_{y_{0}}^{XR} & =\Gamma\frac{e^{-\delta}}{\cosh(\delta)}, & \Gamma_{y_{1}}^{XR}=\Gamma\frac{e^{\delta}}{\cosh(\delta)}.
\end{aligned}
\end{equation}
The dimensionless parameters $\delta$ and $\Delta$ thus allow us
to control the tunneling rates under different conditions. When $\delta=\Delta=0$,
this corresponds to the scenario of completely symmetric and equal
tunneling rates. Conversely, in the limit where $\delta$ approaches
infinity, an electron in channel $X$ can only enter and exit from
reservoir $XL$ at energy $\varepsilon_{X}$, whereas tunneling processes
to reservoir $XR$ are permitted at energy $\varepsilon_{X}+U$. In
this limit, transport is possible only through energy exchange with
channel $Y$. Similarly, the parameter $\Delta$ controls the behavior
of channel $Y$. According to Eqs. (\ref{eq:SRX})-(\ref{eq:IY}),
(\ref{eq:THX}), and (\ref{eq:THY}), the entropy flows associated
with subsystem $X$ and $Y$
\begin{eqnarray}
\dot{S}_{r}^{X} & = & -\beta\Delta\mu_{X}\left(J_{x_{0}x_{1}\mid y_{0}}^{XR}+J_{x_{0}x_{1}\mid y_{1}}^{XR}\right)\\
 &  & -\beta U\left(J_{y_{1}y_{0}\mid x_{0}}^{YL}+J_{y_{1}y_{0}\mid x_{0}}^{YR}\right),\nonumber 
\end{eqnarray}

\begin{eqnarray}
\dot{S}_{r}^{Y} & = & -\beta\Delta\mu_{Y}\left(J_{y_{0}y_{1}\mid x_{0}}^{YR}+J_{y_{0}y_{1}\mid x_{1}}^{YR}\right)\\
 &  & -\beta U\left(J_{x_{1}x_{0}\mid y_{0}}^{XL}+J_{x_{1}x_{0}\mid y_{0}}^{XR}\right),\nonumber 
\end{eqnarray}
the rate of informations

\begin{equation}
\dot{I}^{X}=\left(J_{y_{1}y_{0}\mid x_{0}}^{YL}+J_{y_{1}y_{0}\mid x_{0}}^{YR}\right)\ln\frac{p\left(x_{0},y_{1}\right)p\left(x_{1},y_{0}\right)}{p\left(x_{0},y_{0}\right)p\left(x_{1},y_{1}\right)},
\end{equation}

\begin{equation}
\dot{I}^{Y}=\left(J_{x_{1}x_{0}\mid y_{0}}^{XL}+J_{x_{1}x_{0}\mid y_{0}}^{XR}\right)\ln\frac{p\left(y_{0},x_{1}\right)p\left(y_{1},x_{0}\right)}{p\left(y_{0},x_{0}\right)p\left(y_{1},x_{1}\right)},
\end{equation}
and the coefficients
\begin{equation}
\begin{aligned}\Theta^{X} & =\frac{1}{2}\left(\varepsilon_{X}-\mu_{XL}\right)^{2}\left[W_{10\mid0}^{XL}p(0,0)+W_{01\mid0}^{XL}p(1,0)\right]\\
 & +\frac{1}{2}\left(\varepsilon_{X}-\mu_{XR}\right)^{2}\left[W_{10\mid0}^{XR}p(0,0)+W_{01\mid0}^{XR}p(1,0)\right]\\
 & +\frac{1}{2}\left(\varepsilon_{X}+U-\mu_{XL}\right)^{2}\left[W_{10\mid1}^{XL}p(0,1)+W_{01\mid1}^{XL}p(1,1)\right]\\
 & +\frac{1}{2}\left(\varepsilon_{X}+U-\mu_{XR}\right)^{2}\left[W_{10\mid1}^{XR}p(0,1)+W_{01\mid1}^{XR}p(1,1)\right],
\end{aligned}
\end{equation}

\begin{equation}
\begin{aligned}\Theta^{Y} & =\frac{1}{2}\left(\varepsilon_{Y}-\mu_{YL}\right)^{2}\left[W_{10\mid0}^{YL}p(0,0)+W_{01\mid0}^{YL}p(1,0)\right]\\
 & +\frac{1}{2}\left(\varepsilon_{Y}-\mu_{YR}\right)^{2}\left[W_{10\mid0}^{YR}p(0,0)+W_{01\mid0}^{YR}p(1,0)\right]\\
 & +\frac{1}{2}\left(\varepsilon_{Y}+U-\mu_{YL}\right)^{2}\left[W_{10\mid1}^{YL}p(0,1)+W_{01\mid1}^{YL}p(1,1)\right]\\
 & +\frac{1}{2}\left(\varepsilon_{Y}+U-\mu_{YR}\right)^{2}\left[W_{10\mid1}^{YR}p(0,1)+W_{01\mid1}^{YR}p(1,1)\right].
\end{aligned}
\end{equation}

\section{results and discussion\label{sec:4}}

\begin{figure}
\includegraphics[scale=0.35]{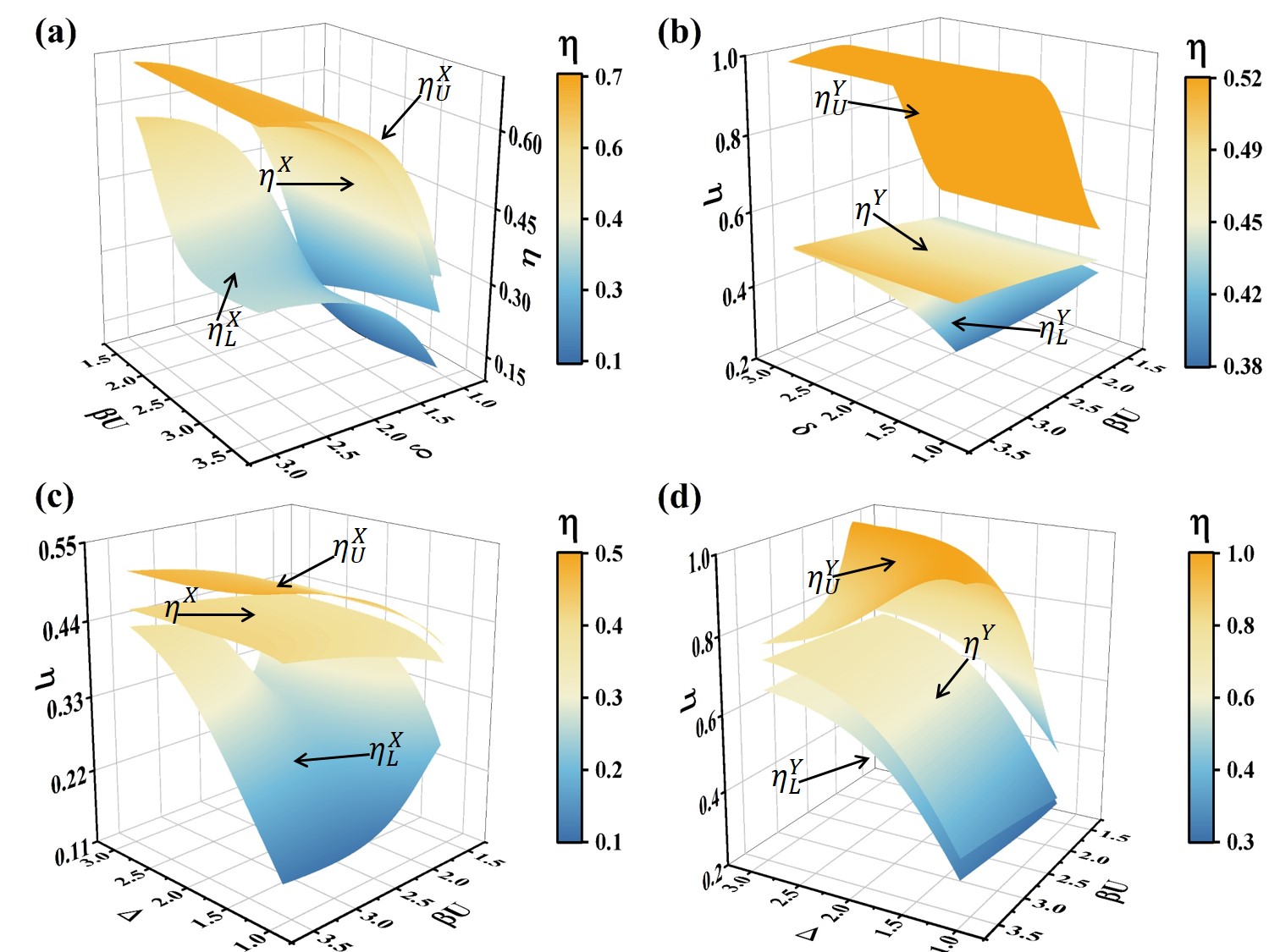}\caption{The efficiencies $\eta^{X}$ and $\eta^{Y}$, upper bounds $\eta_{U}^{X}$
and $\eta_{U}^{Y}$, and lower bounds $\eta_{L}^{X}$ and $\eta_{L}^{Y}$
for the double-quantum-dot system vary with the dimensionless Coulomb
interaction strength $\beta U$ and the parameters $\delta$ and $\Delta$
associated with the tunneling rate, where $\beta\left(\mu_{XR}-\mu_{XL}\right)=1$,
$\beta\left(\mu_{YR}-\mu_{YL}\right)=2,\Gamma=1,\varepsilon_{X}=\frac{\left(\mu_{XR}+\mu_{XL}\right)}{2}-\frac{U}{2}$,
and $\varepsilon_{Y}=\frac{\left(\mu_{YR}+\mu_{YL}\right)}{2}-\frac{U}{2}$.
For (a) and (b), the variations are with $\beta U$ and $\delta$,
where $\Delta=1.5$. For (c) and (d), the variations are with $\beta U$
and $\Delta$, where $\delta=1.5$.}
\end{figure}

Figure 2 plots the efficiencies $\eta^{X}$ and $\eta^{Y}$, upper
bounds $\eta_{U}^{X}$ and $\eta_{U}^{Y}$, and lower bounds $\eta_{L}^{X}$
and $\eta_{L}^{Y}$ for subsystems $X$ and $Y$ varying with the
dimensionless Coulomb interaction strength $\beta U$ and the parameter
$\delta$ and $\Delta$ associated with the tunneling rate. 

From Figure 2(a), it is evident that for subsystem $X$, the discrepancy
between the upper bound $\eta_{U}^{X}$ and the actual efficiency
$\eta^{X}$ is very minimal, and this difference decreases significantly
as the value of $\delta$ increases. Furthermore, when $\delta$ is
held constant, the lower bound $\eta_{L}^{X}$ becomes progressively
tighter as $\beta U$ decreases. It is worth emphasizing that when
both $\beta U$ and $\delta$ are small, the three efficiencies\textemdash upper
bound, actual efficiency, and lower bound\textemdash converge closely
to one another.

On the other hand, Figure 2(b) presents the results for subsystem
$Y$. These results indicate that the upper bound $\eta_{U}^{Y}$
becomes tighter primarily when $\beta U$ is small. As $\beta U$
increases, this upper bound approaches the general result, namely
$\eta^{Y}\leq1$. Therefore, the discussion mainly pertains to the
lower bound $\eta_{L}^{Y}$, which becomes very tight under low limits.
When $\beta U$ is fixed, a larger $\delta$ improves the effectiveness
of the lower bound as a limit. Similarly, when $\delta$ is fixed,
a larger lower limit of $\beta U$ enhances the effectiveness of this
boundary. When both $\beta U$ and $\delta$ are large, the lower
bound is notably stringent.\textbf{ }When we adjust $\beta U$ and
$\Delta$, the results are shown in Figures 2(c) and (d). It can be
observed that the overall trend is consistent with the adjustments
of $\beta U$ and $\delta$. These show that our boundaries are reliable.

Additionally, it is important to note that in the definition of efficiency
in Eq. (21), we assume $\dot{W}^{X}\leq0$ and $\dot{W}^{Y}>0$ for
this model. Therefore, for the quantum dot model described in Sec.
III, there exists a certain range of parameter values that enable
the proper operation of the bipartite system. For example, as shown
in Figure 2(a), when both $\beta U$ and $\delta$ are small, the
three efficiencies converge. This convergence occurs because a small
value of $\beta U$ indicates a very weak interaction between the
two subsystems. Simultaneously, within the range of small values of
$\delta$, the work $\dot{W}^{X}$ tends to zero. This results in
the bipartite system failing to operate normally. Consequently, all
three efficiencies converge towards zero under these conditions.

\section{conclusions\label{sec:5}}

This paper successfully establishes a foundational lower bound on
the entropy production rate of subsystems by employing the Cauchy-Schwarz
inequality, leading to a significant advancement in understanding
the efficiency boundaries of dual subsystems. The derivation of these
efficiency bounds, grounded in the lower limit of entropy production,
ensures applicability to both finely and coarsely-grained systems.
Our empirical investigations, utilizing the double-quantum-dot system,
not only validate the proposed inequalities but also demonstrate their
capability to significantly tighten the efficiency boundaries. This
work, deeply rooted in the principles of stochastic thermodynamics,
offers invaluable insights applicable to a wide range of quantum-dot
systems.
\begin{acknowledgments}
This work has been supported by the National Natural Science Foundation
(Grants No. 12075197 and 12364008), Natural Science Foundation of
Fujian Province (Grant No. 2023J01006), Fundamental Research Fund
for the Central Universities (No. 20720240145), Guiding Project of
Fujian Provincial Department of Science and Technology (2021H0051),
and Talent Project of Quanzhou Science and Technology Bureau (2020C029R). 
\end{acknowledgments}

\appendix

\section{Derivation of the bound of the entropy production rate of subsystem
$X$\label{sec:appendix}}

In this section, we provide a brief derivation of the efficiency limit
for learning. Firstly, we evaluate the upper bound of $\left|\dot{S}_{r}^{X}\right|$
as follows
\begin{widetext}
\begin{equation}
\begin{aligned}\left|\dot{S}_{r}^{X}\right| & =\left|\frac{1}{2}\sum_{x,x^{\prime},y,v}J_{xx^{\prime}\mid y}^{v}\ln\frac{W_{xx^{\prime}\mid y}^{v}}{W_{x^{\prime}x\mid y}^{v}}\right|\\
 & =\left|\frac{1}{2}\sum_{x,x^{\prime},y,v}\ln\frac{W_{xx^{\prime}\mid y}^{v}}{W_{x^{\prime}x\mid y}^{v}}\sqrt{W_{xx^{\prime}\mid y}^{v}p\left(x^{\prime},y\right)+W_{x^{\prime}x\mid y}^{v}p(x,y)}\frac{J_{xx^{\prime}\mid y}^{v}}{\sqrt{W_{xx^{\prime}\mid y}^{v}p\left(x^{\prime},y\right)+W_{x^{\prime}x\mid y}^{v}p(x,y)}}\right|\\
 & \stackrel{(1)}{\leq}\frac{1}{2}\sqrt{\sum_{x,x^{\prime},y,v}\left(\ln\frac{W_{xx^{\prime}\mid y}^{v}}{W_{x^{\prime}x\mid y}^{v}}\right)^{2}\left[W_{xx^{\prime}\mid y}^{v}p\left(x^{\prime},y\right)+W_{x^{\prime}x\mid y}^{v}p(x,y)\right]}\sqrt{\sum_{x,x^{\prime},y,v}\frac{\left(J_{xx^{\prime}\mid y}^{v}\right)^{2}}{W_{xx^{\prime}\mid y}^{v}p\left(x^{\prime},y\right)+W_{x^{\prime}x\mid y}^{v}p(x,y)}}\\
 & \stackrel{(2)}{\leq}\frac{1}{2}\sqrt{\sum_{x,x^{\prime},y,v}\left(\ln\frac{W_{xx^{\prime}\mid y}^{v}}{W_{x^{\prime}x\mid y}^{v}}\right)^{2}\left[W_{xx^{\prime}\mid y}^{v}p\left(x^{\prime},y\right)+W_{x^{\prime}x\mid y}^{v}p(x,y)\right]}\sqrt{\frac{1}{2}\sum_{x,x^{\prime},y,v}J_{xx^{\prime}\mid y}^{v}\ln\frac{W_{xx^{\prime}\mid y}^{v}p\left(x^{\prime},y\right)}{W_{x^{\prime}x\mid y}^{v}p(x,y)}}\\
 & =\frac{1}{2}\sqrt{\sum_{x,x^{\prime},y,v}\left(\ln\frac{W_{xx^{\prime}\mid y}^{v}}{W_{x^{\prime}x\mid y}^{v}}\right)^{2}\left[W_{xx^{\prime}\mid y}^{v}p\left(x^{\prime},y\right)+W_{x^{\prime}x\mid y}^{v}p(x,y)\right]}\sqrt{\dot{\sigma}^{X}}\\
 & =\sqrt{\frac{1}{2}\sum_{x,x^{\prime},y,v}\left(\ln\frac{W_{xx^{\prime}\mid y}^{v}}{W_{x^{\prime}x\mid y}^{v}}\right)^{2}W_{xx^{\prime}\mid y}^{v}p\left(x^{\prime},y\right)\sqrt{\dot{\sigma}^{X}}}\\
 & =\sqrt{\Theta^{X}\dot{\sigma}^{X}}.
\end{aligned}
\label{eq:boundderivation}
\end{equation}
This evaluation incorporates the Cauchy-Schwartz inequality \citep{funo2019speed,shiraishi2018speed}
and the inequality $\frac{(x-y)^{2}}{x+y}\leq\frac{x-y}{2}\log\frac{x}{y}$
for nonnegative $x$ and $y$ in step $(1)$ and $(2)$, respectively.
With the application of Eq. (\ref{eq:boundderivation}), we naturally
derive the lower limit on the entropy-production rate of subsystem
$X$, as presented in Eq. (\ref{eq:epbounds}) in the main text.
\end{widetext}

\bibliographystyle{apsrev4-1}
\bibliography{QD0628}

\end{document}